\documentclass[preprint,12pt]{aastex}

\newcommand{\ea}{{\it et al.}}
\newcommand{\eg}{{\it e.g.}}
\newcommand{\ie}{{\it i.e.}}

\newcommand{\msol}{\rm{M}$_\odot$}
\newcommand{\mdot}{\rm{M}$_\odot$~yr$^{-1}$}
\newcommand{\kms}{km~$\rm{s}^{-1}$}

\newcommand{\beq}{\begin{equation}}
\newcommand{\eeq}{\end{equation}}
\newcommand{\bdm}{\begin{displaymath}}
\newcommand{\edm}{\end{displaymath}}

\newcommand{\jcomp}{\rmfamily{J.~Comp.~Phys.}}
\newcommand{\jfd}{\rmfamily{J.~Fluid~Dynam.}}
\newcommand{\jsc}{\rmfamily{SIAM~J.~Sci.~Comp.}}

\slugcomment{Submitted to the Astrophysical Journal}

\begin{document}

\title{Stellar Outflows Driven by Magnetized Wide-Angle Winds}

\author{T. A. Gardiner$^1$, A. Frank$^1$ and L. Hartmann$^2$}
\affil{$^1$ Dept. of Physics and Astronomy,\\
       University of Rochester, Rochester, NY 14627-0171}
\affil{$^2$ Harvard-Smithsonian Center for Astrophysics,\\
       Cambridge, MA 02138}

\begin{abstract}
We present two-dimensional, cylindrically symmetric simulations of
hydrodynamic and magnetohydrodynamic (MHD) wide-angle winds
interacting with a collapsing environment.  These simulations have
direct relevance to young stellar objects (YSOs). The results may also
be of use in the study of collimated outflows from proto-planetary and
planetary nebulae.  We study a range of wind configurations consistent
with asymptotic MHD wind collimation.  The degree of collimation is
parameterized by the ratio of the wind density at the pole to that of
the equator.  We find that a toroidal magnetic field can have a
significant influence on the resulting outflow, giving rise to a very
dense, jet-like flow in the post-shock region.  The properties of the
flow in this region are similar to the asymptotic state of a
collimated MHD wind.  We conclude that wide-angle MHD winds are quite
likely capable of driving molecular outflows.  Due to difficulty in
treating MHD winds ab-initio in simulations we choose magnetic field
strengths in the wind consistent slow magnetic rotators. While MHD
launched winds will be in the fast rotator regime we discuss how our
results, which rely on toroidal pinch effects, will hold for stronger
field strengths.
\end{abstract}


\section{INTRODUCTION}

\par
Newly formed stars drive copious mass back into the environments which
gave them birth.  Stars at the end of their lives also drive strong
outflows into the interstellar medium. Mass loss from young stellar
objects (YSOs) takes the form of highly collimated Herbig-Haro jets
and molecular outflows \citep{Reipurth97}, while in low mass, evolved
stars, mass loss takes the form of proto-planetary (PPNe) and
planetary nebulae (PNe) \citep{Balick00}. High resolution observations
of PPNe and PNe have revealed outflow features that are very similar
to those found in YSOs \citep{Sahai00}.  In particular, recent visible
and molecular observations of PPNe and young PNe have revealed the
presence of both well collimated jets and broader wind-swept bubbles.
Even though PNe and YSOs are at different extremes of stellar
evolution, the similarity manifested in their mass loss raises
intriguing questions about the processes driving and shaping their
outflows. For example, outflows from PPNe appear to have more momentum
than can be supplied by the central star's luminosity ($L_*/c \ll
\dot{M}V$) \citep{Alcolea00}, a situation which also occurs in YSOs.
Recently, it has been shown that PPNe winds may derive their
acceleration and collimation from the same magnetohydrodynamic (MHD)
forces believed to be at work in YSOs \citep{Blackmanea01}. Thus,
there exits the possibility that a cross fertilization between the two
fields may yield some insights into the origins of magnetized winds
and their dynamical interaction with surrounding media. In this paper
we focus on one specific issue in the study of stellar outflows: the
role of shocked magnetic fields in shaping wind blown bubbles.

\par
Wind blown bubbles have long been part of the theoretical explanation
for PNe.  The interaction of a fast wind from the hot central star of
a PN with a slower, denser, and possibly aspherical AGB wind has been
successful at explaining many properties of spherical PN
\citep{Kwokea77,KahnWest85,Balick87,Icke88,Frankea93,Mellema97}. New
images of PPNe and young PNe, however, show features which can not be
explained with a purely hydrodynamic interacting stellar wind model
\citep{Delamarter2000}. With an eye toward understanding the
intermediate to large scale shaping of PNe, \citet{ChevalierLuo94}
drew attention to the role of shock amplified toroidal magnetic fields
in the wind-blown bubble scenario.  We will refer to this as the
magnetized wind-blown bubble (MWBB) model.  A number of MWBB studies
\citep{RozFranco96,GarciaSegura97,GarciaSegura99} have demonstrated
the model's ability to create a rich variety of structures in the
shocked regions of the flow. A central assumption of the MWBB model,
however, is that the magnetic field strength in the central wind is
sufficiently weak that intrinsic MHD collimation can be
ignored. \citet{Gardiner2001} recently studied the role of intrinsic
collimation in the magnetized central wind showing that for many cases
of interest, the freely expanding wind may experience significant
collimation before it passes through the wind shock. In this paper we
borrow the central idea from \citet{ChevalierLuo94} regarding the
shock amplification of toroidal magnetic fields and use the results of
\citet{Gardiner2001} to explore MHD, wide-angle wind driven outflows
from YSOs.

\par
Wind blown bubbles have also been invoked to explain YSO
outflows. \citet{Shu_91} presented a model wherein a radially directed
wind from a central source sweeps up strongly cooling ambient material
like a snowplow. This {\it snowplow} model was critically examined by
\citet{Masson_Chernin_92} who noted that without extreme choices for
the angular variation of ambient density and wind momentum it is
difficult to match the observed mass versus velocity relation.  They
concluded that the driving force was unlikely to originate from a wide
angle wind, but rather from a collimated, and possibly precessing, jet
\citep{CherninMasson1995}. \citet{Li_Shu1996_collapse}, however, found
that taking the wind parameters from X-wind model \citep{Shu_Xwind_5}
and the ambient density as described by \citet{Li_Shu1996_collapse} an
improved agreement with observations could be found.  More recently,
\citet{Matzner99} have argued that the asymptotic form of the density
and toroidal magnetic field strength in the X-wind model
\citep{Shu_Xwind_5} is generally applicable to all MHD outflows.
Moreover they argue that the power-law mass vs. velocity relations
typically observed for molecular outflows is recovered for a wide
variety of ambient density distributions.

\par
More recently, \citet{Delamarter2000} used time-dependent, purely
hydrodynamic simulations of winds driven into infalling environments,
characteristic of YSOs \citep{Hartmann1996}, to study very young
outflows \citep{Wilkin1999}. It was concluded that the snowplow model
could not embrace nonlinear effects which may yield changes in
momentum distributions in the resultant
outflows. \citet{Delamarter2000} focused on the effect of
shock-focusing, where the wind strikes the bubble shell obliquely and
is redirected to flow along the walls of the shell. The flow of
shocked wind material toward the poles has the effect of substantially
changing the bubble's morphology and evolution from that obtained in a
simple snowplow model. {\it In addition, it was found that inflow ram
pressure produced significant collimation.}  In a different set of
purely hydrodynamic simulations, \citet{Ostriker01} examine both wind
driven and jet driven models, concluding that only some observed
outflows have properties consistent with wide-angle winds.  Their wind
blown bubble simulations do not, however, include infall and their use
of an isothermal equation of state produces what appear to be very
thin bubble shells. It is not clear if shock-focusing could become
manifest in such conditions. Thus, even in pure hydrodynamic models
the applicability of wind blown models remains an open issue.

\par
Magnetic forces are expected to play a central role in launching and
collimating YSO (and perhaps PPN) winds.  Steady state models of MHD
winds have been quite successful at articulating their characteristic
distributions of mass and momentum \citep{Shu_Xwind_5,PudKon00}.
These models rely on a detailed solution of cross-field force balance
via the Grad-Shafranov equation \citep{Shu_91}. Interaction with an
ambient medium will alter these distribution as the wind passes
through the shock bounding the inner edge of the outflow shell. The
subsequent redistribution of mass, momentum, and magnetic energy, as
the plasma seeks equilibrium, will alter downstream (post-shock)
regions of the flow from what might expected via steady MHD wind
models. If the sources are variable, substantial portions of the
outflow cavities may behave differently as well. Thus the
time-dependent interaction of MHD winds with their environments is a
rich subject with important consequences for both YSOs and PNe.

\par
In this paper we describe a set of calculations aimed at elucidating
the interaction of MHD winds from a central source with a collapsing
environment applicable to YSOs.  We show that magnetized wide-angle
winds differ significantly from their hydrodynamic counterparts and
offer a viable explanation for molecular outflows even for slow
magnetic rotators. This paper focuses only on the flow pattern
achieved in the simulations.  In a second paper we plan to discuss
observational consequences of MHD wind driven outflows.  We also note
that due to the difficulty in treating magneto-centrifugally driven
winds ab-initio in larger-scale simulations we choose magnetic field
strengths in the wind consistent slow magnetic rotators. While MHD
launched winds will be in the fast rotator regime our results, relying
on toroidal pinch effects, will hold for stronger field strengths.
This level of abstraction allows us to isolate key effects in the
evolution of magnetized winds which will result regardless of the
field strengths and builds on previous studies which only considered
pure hydrodynamic winds \citep{Delamarter2000,Ostriker01}.

\section{SIMULATION DETAILS}

We present axisymmetric (2.5-D) numerical simulations calculated with
a recently constructed Godunov type scheme for the system of ideal
MHD.  We implement the ``Positive Scheme'' \citep{LiuLax,LaxLiu} which
makes use of a Roe linearized Jacobian matrix, \eg{}
\citet{CargoGallice}.  Optically thin radiative cooling is included by
making use of the Dalgarno-McCray coronal cooling curve in tabulated
form.  Below $10^4$ K the cooling is turned off.  The cooling is
implicitly differenced and solved using Brent's algorithm
\cite{NumRecC}.  The gravitational field of a central source is
included using a first order, explicit Euler step.  The cooling and
gravitational terms are operator split from the MHD integration.  The
system of ideal MHD is integrated in an unsplit (directional
splitting) fashion and second order accuracy is obtained through the
Runge-Kutta method of \citet{ShuOsherRK}.

\par
The simulations presented in this paper use the $f'=10$ simulation of
\citet{Delamarter2000} as a starting point.  The parameter
\beq
f' = \frac{\dot{\rm{M}}_i}{\dot{\rm{M}}_w} ~,
\eeq
measures the ratio of the inflow mass flux to the wind mass flux.
(Note that Table 1 in that paper contains typographical errors such
that the infall mass flux should read $\dot{\rm{M}}_i$ and be in units
of $10^{-7}$ \mdot{} rather than $10^{-6}$ \mdot{}.)  The environment
is derived from simulations of a collapsing, rotating, axially
symmetric sheet \citep{Hartmann1996}.  This study is extended in two
ways. First, a magnetic field, which is believed to be the principal
agent for wind launching and collimation, is introduced into the
wind. Second, an angular variation of the wind density and magnetic
field strength is introduced which is consistent with magnetic
collimation at small radii.

\par
It is straight-forward to show from the $\phi$-component of the
induction equation and the momentum equation that for $r\gg R_A$,
where $r$ is the spherical radius and $R_A$ is the Alfv\'en radius,
\bdm
\frac{B_\phi}{B_r} \approx \frac{-r\Omega\sin(\theta)}{v_r} ~.
\edm
The other parameters in this equation are $\Omega$, the angular
velocity at the base of the wind, and $v_r$, the radial wind
velocity. This expression assumes a radially directed wind and solid
body rotation.  From the expression above we find that the toroidal
field will likely dominate over the poloidal. Thus for $\sin(\theta) =
1/\sqrt{2}$, $v_r = 200$ \kms, $r=13$ AU we find $B_\phi/B_r \sim
6\times10^6 * \Omega$ where $\Omega= 2\pi/\tau$ and $\tau$ is the
rotational period in seconds.  It is not immediately clear which
rotational period should used in these expressions, the stellar
rotational period, or the period of the gas in orbit around the YSO
at a few stellar radii. For example, using the sun's rotational period
of 25 days we find $B_\phi/B_r \sim 20$. The higher rotation rates
associated with YSOs should lead to stronger field ratios than
calculated above and it is likely that $B_\phi/B_r > 100$.  Moreover,
in the presence of cooling the post-shock toroidal field strength can
be amplified by orders of magnitude over it's pre-shock value.  Thus
we begin our calculations by approximating the magnetic field in the
wind as being purely toroidal . The strength of the toroidal magnetic
field is controlled by the parameter $\sigma=({V_m}/{V_\infty})^3$
where $V_m$ is the Michel velocity and $V_\infty$ is the asymptotic
wind velocity \citep{Michel69,BelcherMacGregor76,Gardiner2001}.  In
this study we will be primarily concerned with slow magnetic rotators,
\ie{} $\sigma\ll 1$.  As we will see, since the magnetic field will
come to strongly influence the evolution of the outflow, our results
will be valid for fast rotators (and their stronger initial fields) as
well.

\par
The angular variation of the wind density and magnetic field is
parameterized by $\chi$, the ratio of the density at the pole
$(\theta=0)$ to the density at the equator $(\theta=\pi/2)$ in the
wind.  We refer to this ratio as the pole to equator density contrast
and study the range of $1\le\chi\le 9$.  Given these two parameters,
$\sigma$ and $\chi$, the wind density
\beq
\rho(r,\theta) = \left(\frac{\dot{\rm{M}}}{4\pi r^2 V_\infty}\right)
\left(\frac{\sqrt{\chi}}{\chi\sin^2(\theta) + \cos^2(\theta)}\right) ~,
\eeq
and toroidal magnetic field component
\beq
B_\phi(r,\theta) =
\left( \frac{\sqrt{\sigma \dot{\rm{M}} V_\infty} \sin(\theta)}{r} \right)
\left(\frac{\sqrt{\chi}}{\chi\sin^2(\theta) + \cos^2(\theta)}\right) ~.
\eeq
In these expressions we've assumed that $r$ is sufficiently large that
the radial component of the wind velocity $v_r \approx V_\infty$.  For
$\chi =1$ we obtain the classic split monopole, spherically symmetric
wind conditions.  For $(\chi-1)\sin^2(\theta)\gg 1$ the density and
toroidal field strength take the form $B_\phi \propto 1/\varpi$ and
$\rho \propto 1/\varpi^2$, where $\varpi = r \sin(\theta)$.  This form
is consistent with the asymptotic form of the X-wind solution
\citep{Shu_Xwind_5} and in this way the parameter $\chi$ models the
intrinsic magnetic collimation at small radii.  The parameters $\chi$
and $\sigma$ are not actually independent of one another, but rather
are connected through a self consistent solution at small radii, \ie{}
the launching and collimation process.  However, since a complete
simulation of the wind launching, collimation, and propagation is
beyond the scope of this paper, we will treat them as weakly
independent.

\par
A few details remain to complete the specification of the wind.  All
simulations use a wind with temperature $10^4$ K.  The wind mass loss
rate is $10^{-7}$~\mdot{} and the velocity is taken to be
$200$~\kms{}. The ``wind-sphere'' (the boundary on which the wind
conditions are set) has a radius of $11.25$~AU or 30 grid cells.  The
toroidal magnetic field strength as shown above gives rise to a
current sheet at the equator $(\theta=\pi/2)$.  To smooth out this
singularity, we impose a linear variation on the toroidal magnetic
field within $6$ degrees of the equator.  At the radius of the
wind-sphere this affects approximately 3 grid cells adjacent to the
equatorial plane.  This should not influence the results of the
simulations.  Finally, note that
\beq
\frac{2}{\pi} \int_0^{\pi/2}
\frac{\sqrt{\chi}}{\chi\sin^2(\theta) + \cos^2(\theta)}
~\mathrm{d}\theta = 1 ~.
\eeq
Hence the mass loss rate is independent of the parameter $\chi$.

\section{RESULTS}

\par
Here we focus on the effect of magnetic fields in shaping wide-angle
wind blown bubbles by showing the results of five simulations where
the field strength ($\sigma$) and density stratification ($\chi$)
vary. In figures \ref{fig1} through \ref{fig4} we present the
logarithm of density for the four combinations of $\sigma =0,~0.01$
and $\chi = 1,~9$. Figure \ref{fig5} shows a time sequence of
logarithmic density maps for a more highly stratified model
($\sigma=0.01$, $\chi=100$). Note that $\sigma=0.01$ is approximately
equal to the value for the solar wind, for which $\sigma \approx
0.009$, indicating that this is a slow magnetic rotator, \ie{} the
magnetic field is relatively unimportant in accelerating the wind.  As
we shall see, however, this does not imply that the field is
dynamically unimportant with respect to shaping the bubble
\citep{ChevalierLuo94,GarciaSegura99}. We note also that our
simulations are relevant to the earliest phase of dynamical shaping
since the length and time scales are relatively short ($L < 10^{16}
~cm$, $t < 50~y$). Our explicit goal is to understand the role of
shock amplified magnetic fields and wide-angle winds in shaping the
early evolution of the outflow.

\par
First consider the hydrodynamic models shown in Figure \ref{fig1} and
\ref{fig2} ($\sigma=0$, $\chi = 1,~9$). In the spherical wind
($\chi=1$) case the outflow assumes a characteristic bipolar
morphology. As \citet{Wilkin1999} and \citet{Delamarter2000} have
shown, the aspherical nature of the outflow is the result of both
inertial and infall ram pressure gradients. The morphology is also
effected by the development of shock focusing where the freely flowing
wind impinges the wind shock at an oblique angle.  This leaves high
velocity (potentially supersonic) post-shock gas flowing tangential to
the shock along the walls of the bubble shell. This post-shock
material flows toward the bubble's axis altering its shape and
momentum distribution \citep{Canto1988,Giuliani1982}.  Simplified
calculations by \citet{Delamarter2000} show that mixing of the shocked
ambient and shocked wind material cannot a priori, be expected to
quench shock focusing.  In addition \citet{Delamarter2000} found that
the time-scale for mixing was comparable to the outflow evolution
time-scale meaning the bubble shell was unlikely to be fully
mixed. Thus the shape of the outflow may deviate from what is
predicted by simple snowplow models.

\par
In Figure \ref{fig2} we see the effect of the angular variation in the
wind ram pressure (density stratification in a wide-angle wind,
$\chi=9$).  As has been shown for nebulae surrounding Luminous Blue
Variables, aspherical winds flowing into spherical environments can
produce bipolar bubbles when cooling in the shell is effective
\citep{Frankea1998}.  The higher wind ram pressure at the poles
increases the shock speed at the tip of the bubble, elongating the
outflow. This in turn has a secondary effect of increasing the
obliquity of the wind shock relative to the freely expanding wind.
The degree of shock focusing is thereby increased creating a
converging conical flow at the symmetry axis. Converging conical flows
can be quite effective at producing well collimated jets from the tips
of wind blown bubbles. This effect has been studied in analytical
models \citep{Canto1988}, numerical simulations
\citep{MellemaFrank97}, and more recently, laboratory experiments
\citep{Lebedevea02} which have shown the stability of the collimation
process.

\par
In figures \ref{fig3} and \ref{fig4} we present the logarithm of
density for magnetohydrodynamic wind calculations with $\sigma=0.01$
and $\chi=1,~9$.  Compared with the hydrodynamic calculations, the
magnetic field in the wind has modified the outflow evolution
considerably.  In both of these figures we find a dense, jet-like flow
aligned with the symmetry axis.  These features have been documented
before in spherical wind models appropriate to PPNe
\citep{GarciaSegura99}.  In our simulations we are able to investigate
a new phenomenon, the effect of wind stratification on the subsequent
MWBB evolution.

\par
The origin of axial feature is a simple result of shock amplified
toroidal magnetic fields. The magnetic field is amplified after it
passes through the wind shock.  Strong cooling further compresses the
field, decreasing $\beta = 8\pi P_{gas}/B^2$ and leaving the Lorentz
force out of balance with gas pressure gradients.  When the
magnetosonic wave speed is comparable to or greater than the radial
expansion velocity of the bubble, the plasma in the post-shock region
will be approximately in force balance.  If the radiative cooling
decreases $\beta$ to less than or equal to 1, the Lorentz force will
drive the plasma toward the axis.  Gas pressure near the axis grows
due to the compression driven by the magnetic pinch. While a new
radial force balance will eventually be established, as we discuss
below, the gas and magnetic pressures lead to expansion parallel and
anti-parallel to the $z$-axis. The jets expand along the $z$-axis
leading to an acceleration of the ``head'' of the bubble. When the
wind is spherical, the jet can also push back toward the star
creating a dense ``spike'' that changes the wind shock geometry from
convex to concave. The development of such backflowing jets has been
seen in previous studies \citep{RozFranco96,GarciaSegura99}.

\par
A new feature observed in our simulations is shown in Figure
\ref{fig4} where density stratification in a wide-angle wind produces
increased ram pressure near the axis.  The ram pressure gradient and
corresponding shock focused flow inhibits the backflowing jet. Thus,
for $\chi$ sufficiently large, a density stratified, wide-angle wind
produces a convex wind shock rather than the concave feature seen in
Figure \ref{fig3}.  The quenching of the backflow implies the jets
will not appear to extend back to the source, and one expects hollow
cavities with jets appearing some distance above the source inside the
outflow.

\par
Figure \ref{fig5} shows a time sequence for a highly stratified
wind, $\sigma=0.01$ and $\chi=100$.  In this model the flow is
highly elongated and is quite similar in appearance to simulations
of toroidally dominated MHD jets
\citep{Cerqueira2001,Frankea1998,Frankea2000,Stone00,OSullivan00}.
The long term evolution results in a highly cylindrically
stratified, jet-like outflow. Note that near the base of the flow
the divergent streamlines are still able to carve out a hollow
cavity.

\par
The formation of a steady, dense, well-collimated jet downstream of
the wind shock (as is seen in figures \ref{fig3}, \ref{fig4} and
\ref{fig5}) requires the establishment of a new MHD equilibria. A more
detailed consideration of the simulation results along with simple
results from magneto-statics shows how this occurs \citep{Choudhuri98}.
Establishment of a force balance in a cylindrical plasma column
requires
\beq
\frac{d p_g}{d \varpi} - \frac{\rho v_\phi^2}{\varpi} =
- \frac{1}{\varpi^2}  \frac{d}{d\varpi}  \frac{\varpi^2 B_\phi^2}{8 \pi}
\label{mhdfbal}
\eeq
where the term on the right hand side can be decomposed in to magnetic
pressure and tension components $(\frac{d}{d\varpi} \frac{B_\phi^2} {8
\pi} + \frac{B_\phi^2} {4 \pi \varpi})$.  If the rotation in the wind
is not significant, the behavior of the solution to equation
(\ref{mhdfbal}) is controlled by the parameter $\beta$.  For $\beta \gg
1$, the Lorentz force is insignificant to achieving force balance and
hence we find that the gas pressure is a constant.  For $\beta \ll 1$
we have the opposite situation where the gas pressure is
insignificant.  In this case the magnetic pressure and tension must
balance one another.  From equation (\ref{mhdfbal}) it is apparent
that this force-free configuration occurs for $B_\phi \propto
\varpi^{-1}$. Clearly this approximation cannot hold near the axis
where the gas pressure (or poloidal magnetic field component) must
again become important. From the $z$-component of Ampere's law,
\beq
\frac {1}{\varpi} \frac{d \varpi B_\phi}{d\varpi} =
\frac {4 \pi}{c} j(\varpi)
\label{Ampere_z}
\eeq
we see that this situation describes a jet with a non-zero current
density in the ``core'' surrounded by a current free region.  The
return current flows along the bow shock.

\par
In the simulations we present in this paper, the radial force balance
in the post-shock region is complicated by strong radiative cooling.
In principle, since only the ambient gas is rotating and only the wind
gas is magnetized, the preceding arguments should hold.  We note that
strong cooling in these simulations results in a mixing of the
post-shock gas resulting in some rotation in the bubble plasma.  The
radial equilibrium of the post-shock plasma is then attained through a
combination of the forces present: thermal, centrifugal, and magnetic.

\par
Finally we note that the MHD simulations differ from their
hydrodynamic counterparts in the physics occurring within the
shell.  Note the increased angle between the wind shock and the
slip stream (the contact discontinuity) near the equator in the
MHD models. The critical parameter determining the internal shell
dynamics is the magnetosonic Mach number which is lower than its
sonic counterpart. This implies higher pressures (magnetic + gas)
within the shell. This effect was noted by \citet{Li_Shu1996_disk}
in their study of the interaction of a wide-angle wind with a
flared accretion disk. The relative importance of the magnetic
field in opening the angle between the wind shock and slip stream
is increased in the presence of cooling which depletes the thermal
pressure in the post shock region.

\subsection{MOMENTUM DISTRIBUTION}

Resolution requirements limit our studies to relatively short
length and timescales compared with what is typically observed. In
spite of this limitation we have attempted to characterize one
simple observational characteristic of the outflows in our
simulations. In a future work we provide a more complete
investigation of observational consequences of larger and older
simulated outflows.

In \citet{MassonChernin95} an analysis of the momentum distribution in
a number of molecular outflows was presented.  Averaging in slices
orthogonal to the outflow axis, they found that the momentum
distribution as a function of position along the outflow axis (from
the base to tip of the outflow) is generally peaked in the middle of
the outflow lobe. Comparing their observations to hydrodynamic models
of wind blown bubbles, jet bow shocks and turbulent jet entrainment, 
\citet{MassonChernin95} found that none of the models could recover
the momentum peak in the middle of the outflow flow.

Using a version of our code which can separately track wind and
ambient material we can compare the momentum peak seen in the
observations to what occurs in our models. We have repeated the
calculation of the $\sigma=0.01$, $\chi=1$ model with this new
multiple continuity equation code. In Figure \ref{fig6} we plot
the momentum distribution of the swept up ambient material at 4
different times.  These plots shows two prominent features. One
occurs near the base of the outflow and is ambient material picked
up from the collapsed sheet.  This peak is artificially high due
to a lack of computational resolution across the slip stream. We
note however that even ignoring the resolution issues, the peak at
the base could never account for a significant fraction of the
momentum in a mature (large scale outflow) since momentum
available in a cloud core will be small compared with what has
been swept up from the surrounding inter-core medium. The second
and most important feature is the momentum of the outflow lobe.

At first inspection it might appear that this peak corresponds to the
dense spike (jet) on the axis of the outflow. In fact, this peak does
not occur at the tip of the outflow where the jet elongates the
outflow lobe. Instead the momentum peak is located near the largest
extent in height (z) of what would traditionally be called the ``wind
blown bubble''. Recall that the dense axial jet is formed from wind
material.  Thus it does not contribute at all to the momentum. Note
also that this peak shows a clear tendency to broaden with time 
and coincides with the bubble evolving toward a more prolate
shape.  While it is a large extrapolation to extend these results to
the length scales and evolutionary time scales appropriate to
molecular outflow observations, the results appear suggestive in that
the observed momentum distributions could be recovered if we were to
let these models evolve for longer times allowing the peak to continue
to broaden and smooth.

Perhaps the most promising aspect of our models is the suggestion
of an observational split between a narrow jet which forms from
wind material on the axis (due to axial magnetic forces) and the
broader outflow formed by the rest of the wide-angle wind sweeping
up ambient material.  Thus the \citet{MassonChernin95} finding may
point to a wide-angle wind with perhaps most of the ram pressure
per unit area along the axis, but the bulk of the outward momentum
at some fairly wide opening angle. By creating a jet within the
wind, the addition of magnetic fields to wide angle wind models may
provide the needed physics to overcome the obstacles to finding an
observationally consistent model of molecular outflows
\citep{Cabritea97}.

\section{CONCLUSIONS}

\par
A dichotomy exists in the theoretical picture of molecular outflow
formation.  It appears that, to some degree, wide-angle winds were
abandoned in favor of jets as the driving source for molecular
outflows \citep{Masson_Chernin_92}.  Such conclusions have been, to a
large extent, based upon an analysis of snowplow models
\citep{Shu_91} in which ambient molecular material is driven in
a purely radial direction by a central, wide-angle wind.  In the wake
of these conclusions, it was hoped that both Herbig Haro (HH) jets and
molecular outflows could be unified into a single model for outflows
from YSOs.  More recent studies have shown that wind-angle winds can
be effective in producing molecular outflows when non-linear flow
effects or appropriate wind/ambient conditions are considered
\citep{Wilkin1999,Delamarter2000,Ostriker01}. Since both HH jets and
wide-angle winds are believed to be launched and collimated by
magnetic fields, attempts to unify jets and molecular outflows into a
single consistent phenomenon need to address the role of {\it magnetic
forces and shocks} in shaping the flows.  In this paper we have sought
to address this question.

\par
As we have shown, MHD wide-angle winds with little or no collimation
at small scales give rise to a dense, jet-like, collimated
outflows. We have shown that this is true even for slow magnetic
rotators. Increasing either the collimation at small radii,
parameterized by $\chi$, or the magnetic field strength, parameterized
by $\sigma$, toward the fast magnetic rotator limit will only enhance
these results. As shown previously
\citep{Li_Shu1996_collapse,Matzner99}, steady wide-angle winds with
cylindrical density stratification, (consistent with MHD wind
collimation), are capable of reproducing the observed characteristics
of molecular outflows.  In this paper we have described unsteady
phenomena which result in highly collimated outflows even if there is
little or no collimation to the driving wind. Extrapolating these
simulations to larger radii, one can envision that as the flow
evolves, an evacuated and well collimated channel is created. At
larger radii, the intrinsic MHD collimation will narrow the central
wind, further realizing the results of \citet{Li_Shu1996_collapse} and
\citet{Matzner99}. Thus we conclude that MHD wide-angle winds are
capable of driving molecular outflows.

\par
Our results are also of relevance to models of PNe and PPNe.
\citet{ChevalierLuo94}, \citet{RozFranco96} and \citet{GarciaSegura99}
pioneered the study of the MWBB model in evolved, stellar outflows. As
\citet{Gardiner2001} have shown, however, even winds from slow
rotators are likely to experience collimation before they are
shocked via interaction with the environment. This
conclusion implies that MWBB models should begin with density
stratified wide-angle winds as initial conditions.  The
simulations presented in this paper show that when such wide-angle
winds are included the resultant morphologies differ in
significant ways from those obtained with spherically symmetric
winds.  In particular, the backflows obtained with spherical winds
are suppressed when density stratified flows are included.

\par
Future directions for this line of research include modeling both
the smaller, collimation scales and larger, propagation scales. In
addition, one might question the stability of such thin, dense
magnetized plasma configurations seen in the axial jet.
\citet{GarciaSegura97} studied the three-dimensional evolution of
a magnetized wind for the case of PPNe at low resolution.  That
investigation found that while the dense, jet-like core does suffer
some instabilities its gross morphology was not disrupted giving rise
to similar outflow characteristics seen in two-dimensional
cylindrically symmetric calculations (\ie{} the formation of
streamlined nose cones \citep{Frankea1998}). We note however,
\citet{Cerqueira2001} study jet propagation in three-dimensional jets
and contrary to two-dimensional simulations do not observe the
formation of ``nose-cones''.

\acknowledgements

We wish to thank Chris Matzner for helpful discussions and for
providing us with information on generic wind configurations. This
work was supported by NSF Grant AST-0978765, NASA Grant NAG5-8428 and
the University of Rochester's Laboratory for Laser Energetics.


%
\begin{figure}[!ht]
\caption{Log(density) of purely hydrodynamic spherical wind model.
($\sigma=0$, $\chi=1$), t = 65 y. Spatial scale of grid is
$600 \times 600$ AU.}
\label{fig1}
\end{figure}
\begin{figure}[!ht]
\caption{Log(density) of purely hydrodynamic density stratified
{\it wide-angle}
 wind model $\sigma=0$, $\chi=9$, t = 36 y.}
\label{fig2}
\end{figure}
\begin{figure}[!ht]
\caption{Log(density) of magnetized spherical wind model
$\sigma=0.01$, $\chi=1$, t = 34 y. Note the concave shape of the
wind shock produced by the backflow from the axial jet in the post
shock region. }
\label{fig3}
\end{figure}
\begin{figure}[!ht]
\caption{Log(density) of magnetized density stratified {\it
wide-angle} wind model $\sigma=0.01$, $\chi=9$, t = 26 y.  Note
the shape of the wind shock in now convex.}
\label{fig4}
\end{figure}
\begin{figure}[!ht]
\begin{center}
\end{center}
\caption{Log(density) of magnetized density stratified {\it
wide-angle} wind model $\sigma=0.01$, $\chi=100$, t = 6.5, 13,
19.5, y.  Images measure $150 \times 600$ AU.}
\label{fig5}
\end{figure}
\begin{figure}[!ht]
\begin{center}
\plotone{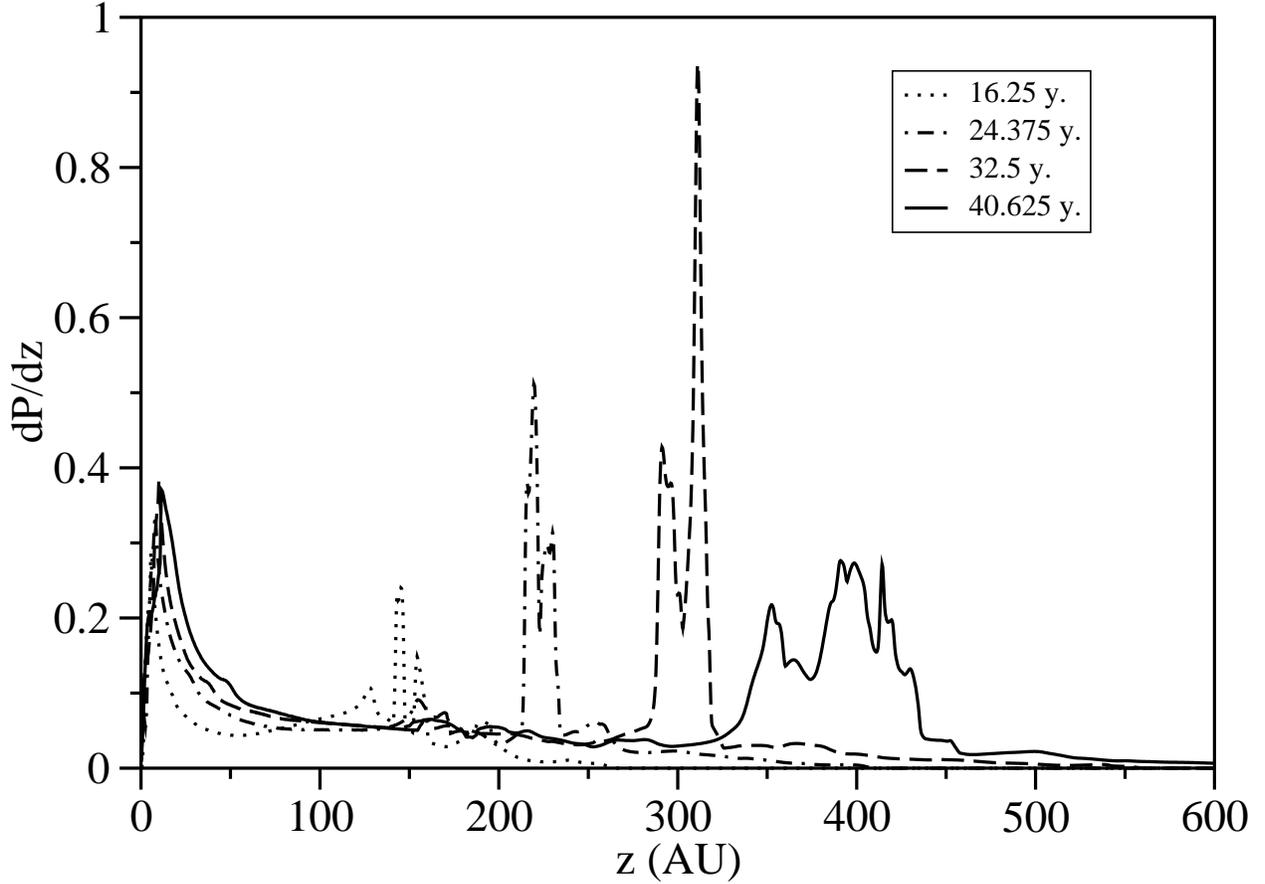}
\end{center}
\caption{ Momentum distribution in swept up ambient material as a
function of position along the outflow axis for the $\sigma=0.01$,
$\chi=1$ model at four times. Plot made by averaging in slices
orthogonal to the outflow axis. The units of the momentum distribution
are \msol{} km s$^{-1}$ pc$^{-1}$.  In this plot the tip of the
outflow occurs at 260, 410, 560, and $>600$ AU at an evolutionary time
of 16, 24, 32, and 40 y respectively.  Also at these times, the apex
of the traditional wind blown bubble is located at 150, 225, 310, and
425 AU respectively.  Thus the momentum peak at large values of $z$
does not correspond to the tip of the outflow where the dense jet
elongates the lobe but is associated swept up ambient material near
the bubble apex.}
\label{fig6}
\end{figure}

\end{document}